\documentclass{article}
\usepackage{arxiv}

\usepackage[utf8]{inputenc} 
\usepackage[T1]{fontenc}    
\usepackage{hyperref}       
\usepackage{url}            
\usepackage{booktabs}       
\usepackage{amsfonts}       
\usepackage{nicefrac}       
\usepackage{microtype}      
\usepackage{lipsum}		
\usepackage{graphicx}
\usepackage[square,sort,comma,numbers]{natbib}
\usepackage{doi}
\usepackage{xcolor}

\title{TJDR: A High-Quality Diabetic Retinopathy Pixel-Level Annotation Dataset}

\author{
Jingxin Mao$^1$\\
\texttt{mjx@tongji.edu.cn}\\
\And
Xiaoyu Ma$^2$\\
\texttt{2031252@tongji.edu.cn}\\
\And
Yanlong Bi$^{2,3*}$\\
\texttt{biyanlong@tongji.edu.cn}\\
\And
Rongqing Zhang$^{1*}$\\
\texttt{rongqingz@tongji.edu.cn}\\
\\
$^1$ School of Software Engineering, Tongji University, Shanghai, China\\
$^2$ Department of Ophthalmology, Tongji Hospital, School of Medicine, Tongji University, Shanghai, China\\
$^3$ Tongji Eye Institute, School of Medicine, Tongji University, Shanghai, China\\
$^*$ Corresponding Author
}

\date{}

\renewcommand{\headeright}{}
\renewcommand{\undertitle}{}
\renewcommand{\shorttitle}{}

\hypersetup{
pdftitle={TJDR: A High-Quality Diabetic Retinopathy Pixel-Level Annotation Dataset},
pdfauthor={Jingxin Mao, Xiaoyu Ma, Yanlong Bi, and Rongqing Zhang},
}

\begin{document}
\maketitle

\begin{abstract}
Diabetic retinopathy (DR), as a debilitating ocular complication, necessitates prompt intervention and treatment. 
Despite the effectiveness of artificial intelligence in aiding DR grading, the progression of research toward enhancing the interpretability of DR grading through precise lesion segmentation faces a severe hindrance due to the scarcity of pixel-level annotated DR datasets.
To mitigate this, this paper presents and delineates TJDR, a high-quality DR pixel-level annotation dataset, which comprises 561 color fundus images sourced from the Tongji Hospital Affiliated to Tongji University.
These images are captured using diverse fundus cameras including Topcon's TRC-50DX and Zeiss CLARUS 500, exhibit high resolution. For the sake of adhering strictly to principles of data privacy, the private information of images is meticulously removed while ensuring clarity in displaying anatomical structures such as the optic disc, retinal blood vessels, and macular fovea.
The DR lesions are annotated using the Labelme tool, encompassing four prevalent DR lesions: Hard Exudates (EX), Hemorrhages (HE), Microaneurysms (MA), and Soft Exudates (SE), labeled respectively from 1 to 4, with 0 representing the background.
Significantly, experienced ophthalmologists conduct the annotation work with rigorous quality assurance, culminating in the construction of this dataset. This dataset has been partitioned into training and testing sets and publicly released to contribute to advancements in the DR lesion segmentation research community.
\end{abstract}

\section{Data Source}
Color fundus images of diabetic patients in the ophthalmology outpatient department of \textbf{Tongji Hospital Affiliated to Tongji University (referred to as Tongji Hospital)}, were retrospectively gathered, resulting in a total of 561 fundus images.
Out of these, 257 images with a resolution of 2,048$\times$2,048 px were captured by the TRC-50DX fundus camera, manufactured by Topcon Corporation in Japan, offering a field of view between 35 to 50 degrees. Additionally, 304 images with 3,912$\times$3,912 px resolution were acquired using the new-generation Zeiss CLARUS 500 True color, HD and ultra-wide-angle fundus camera. This camera represents the latest advancement in fundus photography, as a single image can capture a range of 133°, far surpassing the 45° range of traditional cameras. It provides clear and precise images from the macula to the peripheral retina, enabling more comprehensive detection of peripheral retinal lesions and providing more detailed information on the lesions.
\textbf{During the data collection process, we rigorously adhered to the principles of data privacy protection and secured informed consent. To uphold privacy standards, all identifiable personal information was meticulously removed from the images.}
The images were required to exhibit clear visibility of anatomical structures such as the optic disc, retinal blood vessels, and macular fovea. For images displaying diabetic retinopathy lesions, lesions should be as clear and easy to label as possible. Table~\ref{tab:data_source} furnishes comprehensive details regarding the color fundus images acquired from 2018 to 2022.

\begin{table}[!ht]
\centering
\caption{Data source}
\label{tab:data_source}
\resizebox{0.7\linewidth}{!}{
\begin{tabular}{cccc}
\toprule
Hospital & Number of Photos & Collection Period & Camera Model \\
\midrule
  & 65 & 2018.1--2018.12 & TRC-50DX, Topcon \\
Tongji Hospital  & 103 & 2019.1--2019.12 & TRC-50DX, Topcon \\
Affiliated to  & 89 & 2020.1--2020.12 & TRC-50DX, Topcon \\
Tongji University & 196 & 2021.1--2021.12 & Zeiss CLARUS 500 \\
  & 108 & 2022.1--2022.12 & Zeiss CLARUS 500 \\
\bottomrule
\end{tabular}%
}
\end{table}

\section{Lesion Annotation}
\subsection{Annotation Tool}
\begin{figure}[!ht]
\centering
\resizebox{0.7\linewidth}{!}{
\includegraphics{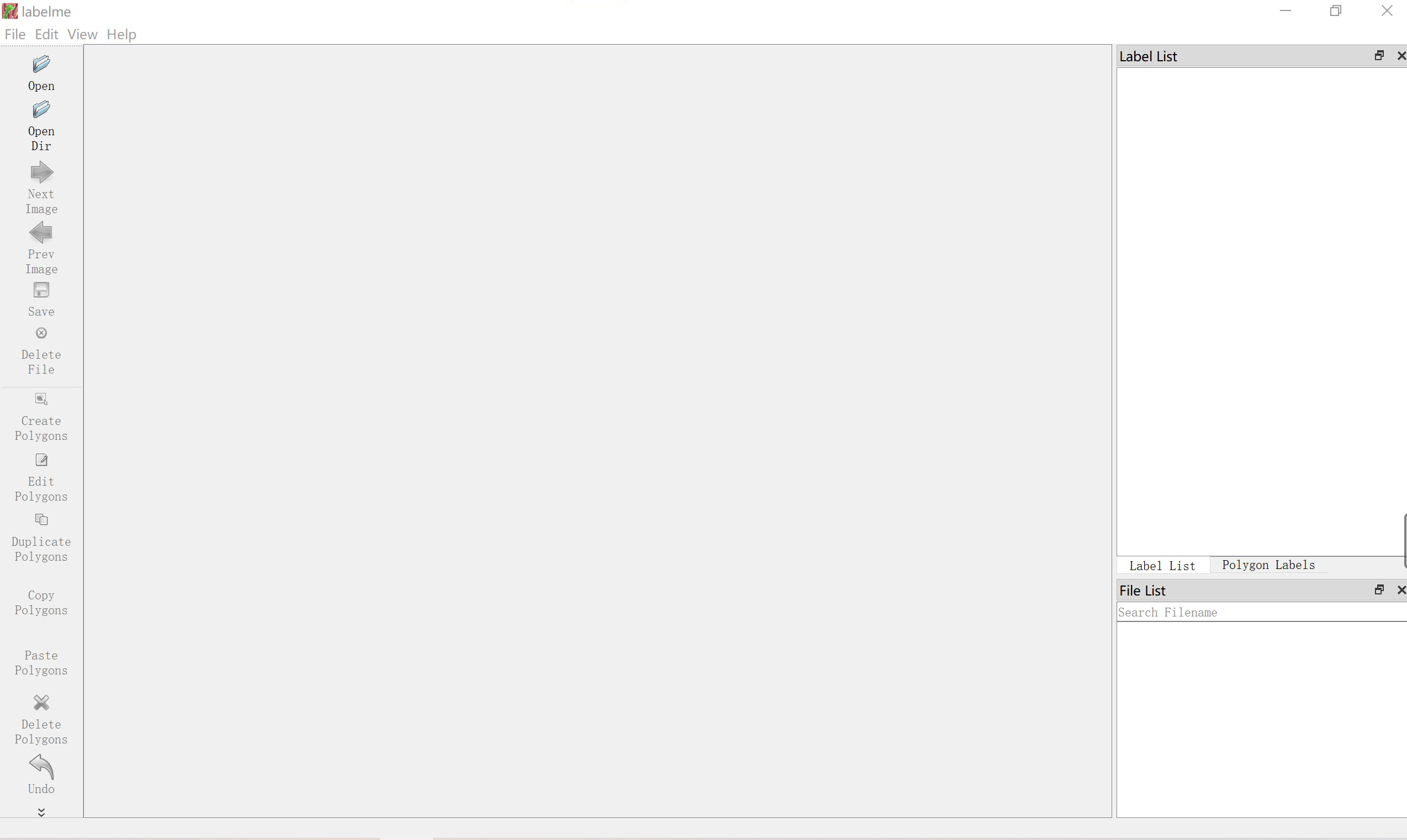}
}
\caption{The main interface of Labelme.}
\label{fig:labelme}
\end{figure}

The annotation tool adopted in this study is Labelme (version: 5.1.1)\footnote{https://github.com/wkentaro/labelme/releases/tag/v5.1.1}, which can create polygons, circles, lines, points, and other forms of borders to meet various types of lesion labeling, and the operation is relatively simple. Using this tool, lesions in pictures can be labeled at the pixel level. Figure~\ref{fig:labelme} shows the main interface of Labelme.

\subsection{Manual Annotation of Single Lesion Candidate Area}
\begin{figure}[!ht]
\centering
\setlength{\abovecaptionskip}{-0.cm}
\setlength{\belowcaptionskip}{-0.4cm}
\resizebox{0.8\linewidth}{!}{
\includegraphics{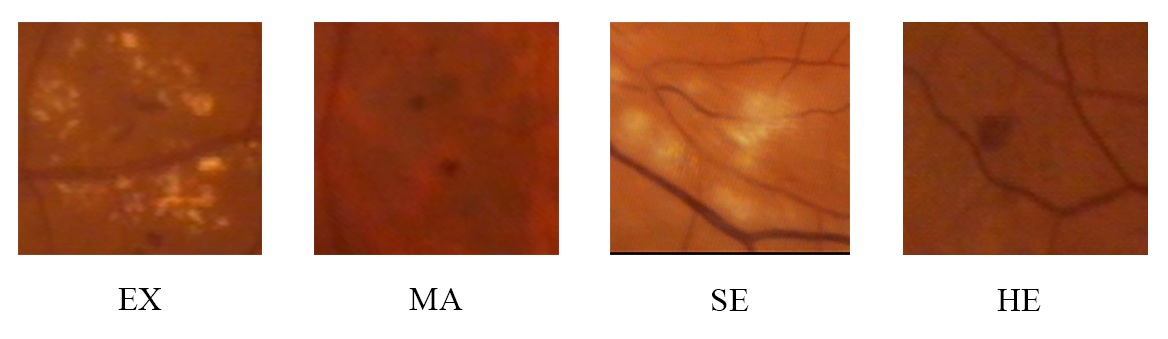}
}
\caption{Illustration of four DR lesion types.}
\label{fig:dr_lesion}
\end{figure}

Upon acquiring the dataset, the annotation process for the identification of diabetic retinopathy (DR) lesions in the images commenced. Common DR lesions encompass Hard Exudates (EX), Hemorrhages (HE), Microaneurysms (MA), Soft Exudates (SE), Neovascularization, and Venous Beading, among others. Considering the variety and complexity of lesions within the dataset, we selected EX, HE, MA, and SE as the lesion labels for annotation (Figure~\ref{fig:dr_lesion} provides an illustration of these four lesion types).

\begin{itemize}

\item Hard Exudates (EX): EX lesions manifest as yellowish, punctate, or patchy areas with clear boundaries and may appear as single entities or in clusters.

\item  Hemorrhages (HE): In the early stages, HE lesions are typically located in the deep layers of the retina, presenting as circular or small dot-like deposits. As the condition progresses, they may develop into superficial streaks or flame-shaped hemorrhages, and in some cases, even large subinternal limiting membrane or subretinal hemorrhages. Small hemorrhages may bear resemblance to MA in terms of color and morphology, making them challenging to differentiate.

\item  Microaneurysms (MA): These appear in retinal images as well-defined, smooth-edged, red, or dark red spots of varying sizes.

\item Soft Exudates (SE): SE lesions exhibit an irregular, ill-defined, cotton-wool-like, grayish-white appearance in retinal images.

\end{itemize}

\begin{figure}[!ht]
\setlength{\abovecaptionskip}{-0.cm}
\setlength{\belowcaptionskip}{0.1cm}
\centering
\resizebox{0.8\linewidth}{!}{
\includegraphics{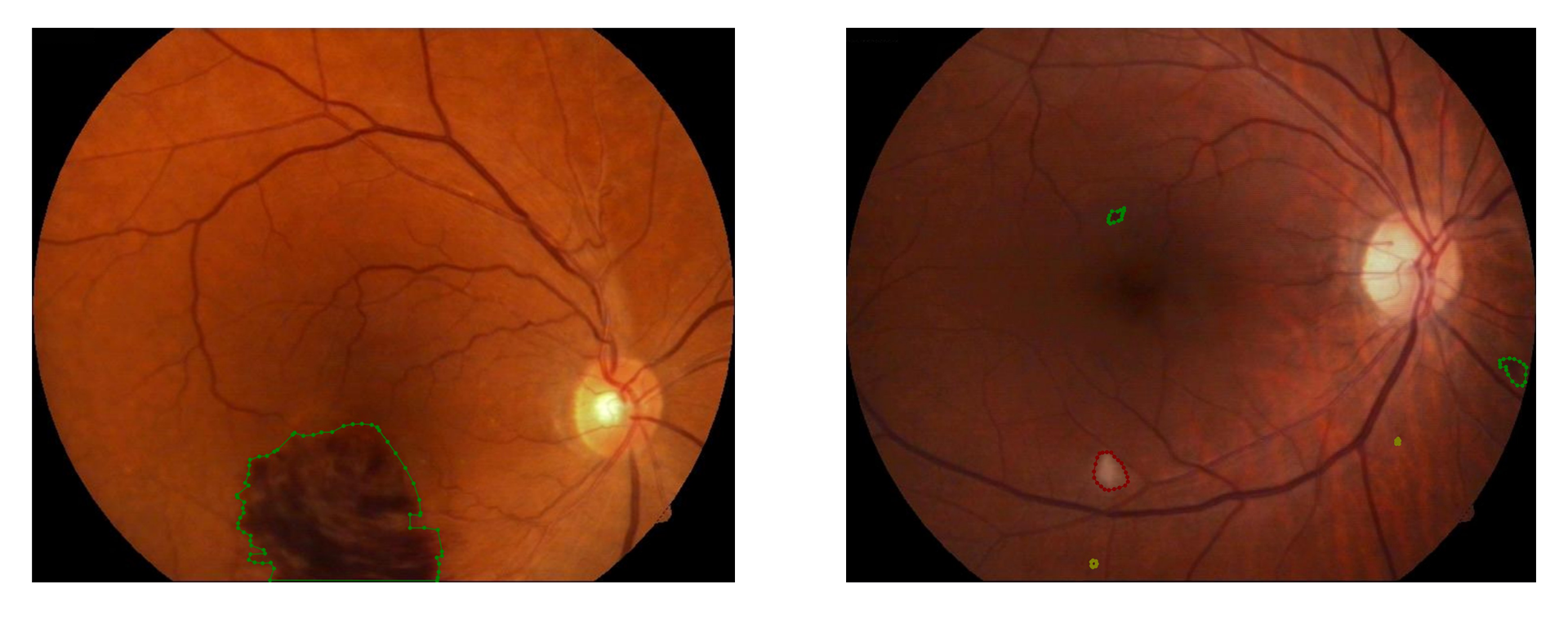}
}
\caption{Example of lesion annotations using Labelme.}
\label{fig:lesion}
\end{figure}

The annotation process involved initial annotations conducted by two experienced ophthalmologists, who independently marked the lesions while being blinded to each other's annotations. Regions where their annotations matched were considered the final annotation results, while areas with discordant annotations were referred to a third, more senior ophthalmologist to determine the ultimate annotations. Subsequently, a random subset of images was subjected to quality assurance by a senior ophthalmologist to ensure the overall annotation quality. Figure~\ref{fig:lesion} provides an example of the annotation process, with each of the four lesion types represented using distinct colors.

\subsection{Dataset Construction}
\begin{figure}[!ht]
\centering
\resizebox{0.4\linewidth}{!}{
\includegraphics{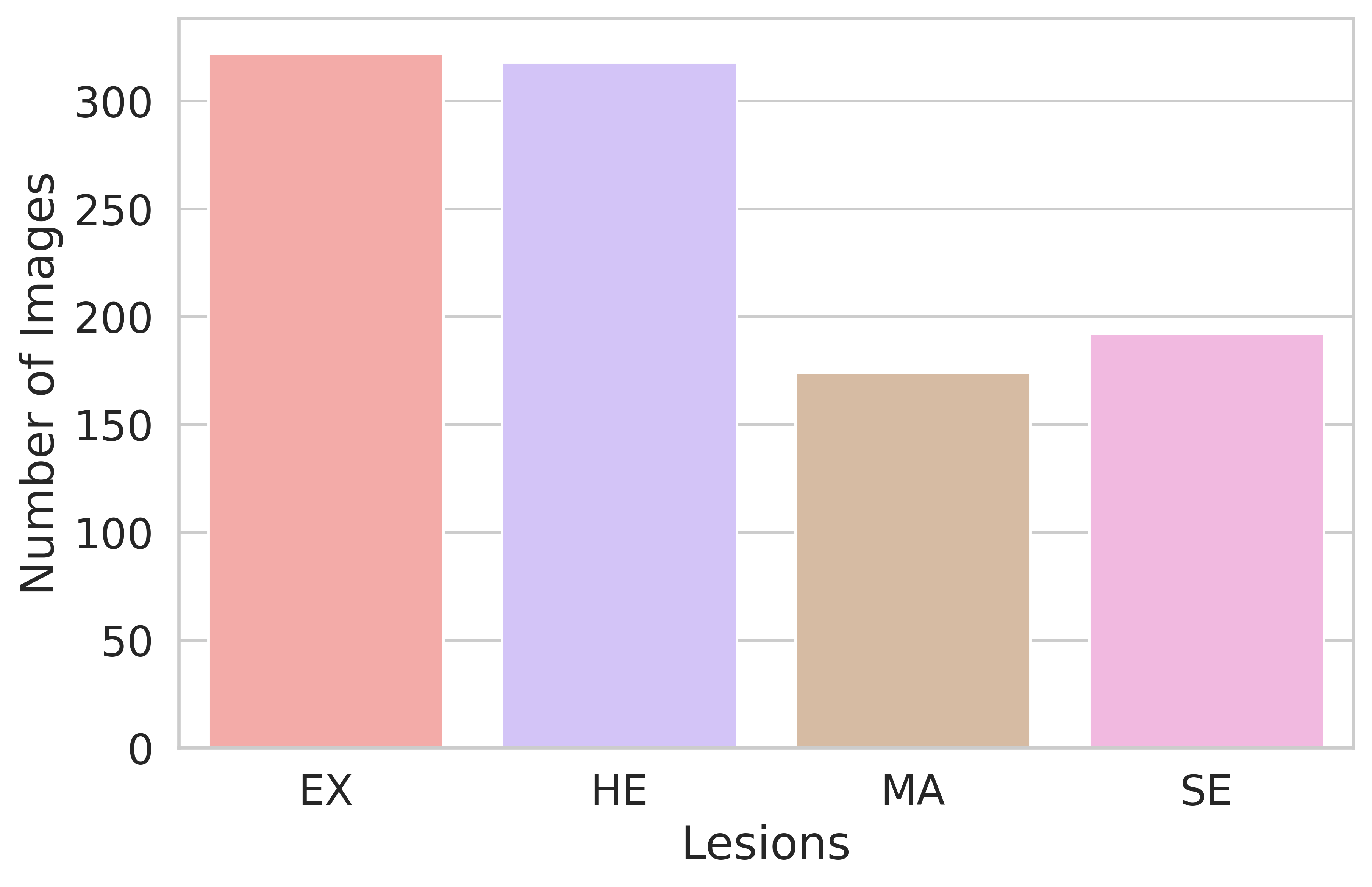}
}
\caption{The number of images annotated with four lesions within the TJDR.}
\label{fig:img_tjdr}
\end{figure}

\begin{figure}[!ht]
\centering
\resizebox{0.3\linewidth}{!}{
\includegraphics{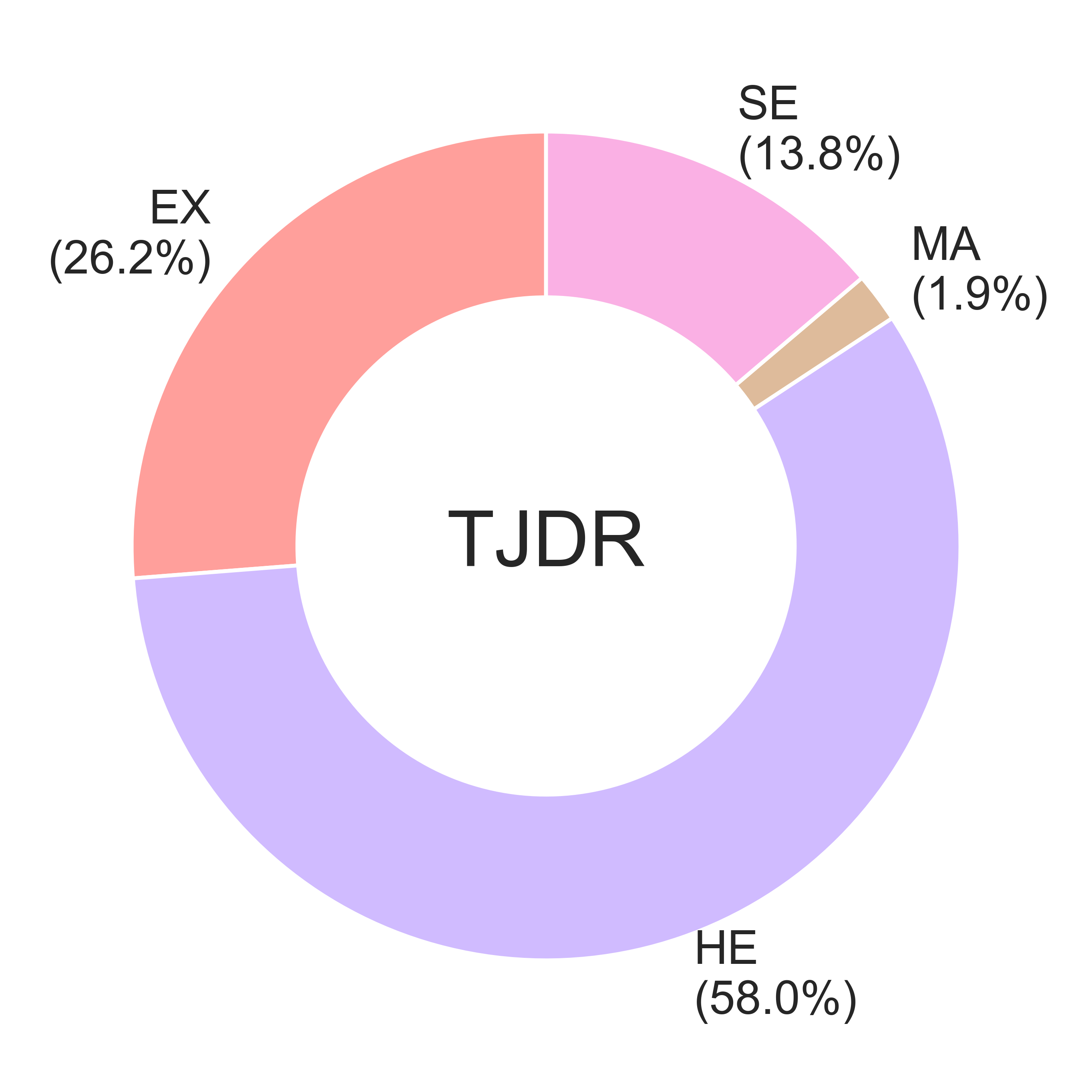}
}
\caption{The proportion of annotated pixels among four lesions in the TJDR.}
\label{fig:pixel_tjdr}
\end{figure}

After meticulous annotation, we have curated a high-quality, fine-grained, pixel-level Diabetic Retinopathy (DR) lesion dataset, named TJDR. Within the TJDR dataset, every image is meticulously annotated at the pixel level, encompassing at least 
one type of lesion among the meticulously categorized EX(1), HE(2), MA(3), and SE(4), where 0 represents the background.
Figure~\ref{fig:img_tjdr} illustrates the number of images in the TJDR dataset corresponding to their respective lesion annotations. It is noticeable that our dataset includes a relatively higher number of images annotated with EX and HE, whereas images annotated with the other two types of lesions are comparatively fewer. Figure~\ref{fig:pixel_tjdr} demonstrates the proportion of annotated pixels for each lesion type within the TJDR dataset, revealing that the combined pixel counts of EX and HE annotations constitute a significant portion of the total annotated lesion pixels (i.e., exceeding 80\%), which draws parallels to the characteristics observed in current pixel-level DR datasets~\cite{porwal2018indian,li2019diagnostic}. Concurrently, annotations for SE and MA display relatively lower pixel counts. Overall, our dataset exhibits higher and more consistent resolution, maintaining a high standard of annotation quality through meticulous annotation procedures. Furthermore, we have partitioned the dataset into training and testing sets in a 4:1 ratio, comprising 448 and 113 images, respectively.

\section{Hospital Qualification}
Tongji Hospital Affiliated to Tongji University (referred to as Tongji Hospital), is a comprehensive university-affiliated hospital and a Grade III Class A hospital. It possesses strong medical and technical capabilities, as well as teaching and research expertise. The Department of Ophthalmology is equipped with central laminar flow purification facilities, including a Zeiss CLARUS 500 ultra-widefield fundus camera from Germany, a CIRRUS OCT 5000 ANGIO device, multiple slit lamps, a TOPCON fundus camera, an ALCON Constellation phacoemulsification machine, and a vitrectomy system. With an adept healthcare team, state-of-the-art equipment, and skilled personnel, the department provides excellent support for the execution of this research project.
The Ophthalmology Ward at Tongji Hospital accommodates more than 30 fixed beds and handles over 200 outpatient visits daily. Outpatients are regularly followed up, exhibit good compliance, and present a diverse range of eye conditions.

\section{Open Access}
We publicly release this dataset, enabling direct downloads without any application process, in hopes of significantly advancing the current research within the DR lesion segmentation community. For access to the download link, please visit \href{https://github.com/NekoPii/TJDR}{\textbf{\textcolor{red}{here}}}.

\section*{Acknowledgment}
This work was supported by the Shanghai Municipal Science and Technology Major Project (2021SHZDZX0100) and the Fundamental Research Funds for the Central Universities.

\bibliographystyle{unsrtnat}

\end{document}